\documentclass[12pt]{article}
\usepackage{physics}
\emergencystretch=\maxdimen
\UseRawInputEncoding
\usepackage[numbers,round,sort&compress]{natbib}

\newcommand\hrefBibPDF[3][]{}

\usepackage{siunitx}
\usepackage{times}
\usepackage{graphicx}

\usepackage[english]{babel}
\usepackage{hyperref}
\hypersetup{pdfstartview={FitH},pdfpagemode={UseNone},
            colorlinks,linkcolor=blue, citecolor=blue, urlcolor=blue,
            bookmarksopen=true, pdfnewwindow=true}
\usepackage[all]{hypcap}

\usepackage{braket}
\newcommand\REMOVE[1]{}

\makeatletter
\let\origcitation\citation
\AtEndDocument{\def\mycites{}%
  \def\citation#1{\g@addto@macro\mycites{\string\nocite{#1}^^J}\origcitation{#1}}}
\AtVeryEndDocument{\newwrite\citeout\immediate\openout\citeout=\jobname.cit_main
  \immediate\write\citeout{\mycites}\immediate\closeout\citeout}
\makeatother

\newenvironment{sciabstract}{%
\begin{quote} \bf}
{\end{quote}}

\title{Generation of tunable quantum entanglement via nonlinearity symmetry breaking in semiconductor metasurfaces} 

\author{Jinyong Ma$^{1,\ast\dagger}$, Tongmiao Fan$^{1\ast}$, Tuomas Haggren$^{1\ast}$, Laura Valencia Molina$^{1,2}$, \\Matthew Parry$^{1}$, Saniya Shinde$^{2,1}$,
Jihua Zhang$^{1,3}$, Rocio Camacho Morales$^{1}$,\\
Frank Setzpfandt$^{2,4}$, 
Hark Hoe Tan$^{1}$, 
Chennupati Jagadish$^{1}$,\\
Dragomir N. Neshev$^{1}$, Andrey A. Sukhorukov$^{1\dagger}$\\
\\
\normalsize{$^{1}$ARC Centre of Excellence for Transformative Meta-Optical Systems (TMOS),}\\
\normalsize{Department of Electronic Materials Engineering, Research School of Physics,}\\
\normalsize{The~Australian National University, Canberra, ACT 2601, Australia}\\
\normalsize{$^{2}$Abbe Center of Photonics, Friedrich Schiller University Jena, Jena, Germany}\\
\normalsize{$^{3}$Songshan Lake Materials Laboratory, Dongguan, 523808, P. R. China}\\
\normalsize{$^{4}$Fraunhofer Institute for Applied Optics and Precision Engineering IOF, Jena, Germany}\\
\\
\normalsize{$^\ast$Co-first authors with equal contribution}\\
\normalsize{$^\dagger$To whom correspondence should be addressed:}\\
\normalsize{E-mail:  \href{mailto:jinyong.ma@anu.edu.au}{jinyong.ma@anu.edu.au} 
E-mail:  \href{mailto:andrey.sukhorukov@anu.edu.au}{andrey.sukhorukov@anu.edu.au}
} 
}

\date{}

\begin{document} 
\sloppy 

\baselineskip=24pt


\maketitle


\begin{sciabstract}
{\em 125-character summary:}
\\\vspace*{0mm}

{\em Abstract:}\\
Tunable biphoton quantum entanglement generated from nonlinear processes is highly desirable for cutting-edge quantum technologies, yet its tunability is substantially constrained by the symmetry of material nonlinear tensors. Here, we overcome this constraint by introducing symmetry-breaking in nonlinear polarization to generate optically tunable biphoton entanglement at picosecond speeds. Asymmetric optical responses have made breakthroughs in classical applications like non-reciprocal light transmission. We now experimentally demonstrate the nonlinear asymmetry response for biphoton entanglement using a semiconductor metasurface incorporating [110] InGaP nano-resonators with structural asymmetry. We realize continuous tuning of polarization entanglement from near-unentangled states to a Bell state. This tunability can also extend to produce tailored hyperentanglement. Furthermore, our nanoscale entanglement source features an ultra-high coincidence-to-accidental ratio of $\approx7\times10^4$, outperforming existing semiconductor flat optics by two orders of magnitude. Introducing asymmetric nonlinear response in quantum metasurfaces opens new directions for tailoring on-demand quantum states and beyond.

\end{sciabstract}


\section*{Introduction}
Tunable photonic quantum entanglement promises to advance information processing beyond classical approaches, facilitating applications in quantum communications~\cite{liaoSatellitetoground2017}, imaging~\cite{ biancofioreQuantum2011c} and computing~\cite{slussarenkoPhotonic2019}. A practical approach to preparing quantum entanglement at room temperature is to engineer photon-pair generation 
through spontaneous parametric down-conversion (SPDC). 
Nonlinear flat-optics devices, incorporating sub-wavelength thick nonlinear materials, are emerging as new platforms for generating photon pairs at the nanoscale~\cite{schulzRoadmap2024a, kuznetsovRoadmap2024, maEngineering2024, guoUltrathin2023, santiago-cruzResonant2022}. In particular, nanostructured metasurfaces supporting optical resonances can enhance and tailor SPDC processes, allowing for the generation of spectral~\cite{santiago-cruzResonant2022}, polarization~\cite{maPolarization2023a, jiaEntangled}, and spatial~\cite{zhangSpatially2022,ma2024quantum} entanglement. However, the tunability of polarization entanglement~\cite{sultanovTunable2024} is subject to the symmetry of nonlinear tensors of materials, including dielectric materials~\cite{maPolarization2023a}, III-V semiconductors~\cite{sultanovFlatoptics2022}, or van der Waals semiconductors~\cite{weissflogTunable2024, fengPolarizationentangled2024}, where the pump photon polarization serves as the only tuning parameter.
As a result, achieving entanglement with comprehensive, real-time, and ultra-fast tunability, which is critical for quantum information processing~\cite{donohueUltrafast2014,kawasaki2024real}, remained a significant challenge~\cite{nichollsUltrafast2017}.


Optical systems with asymmetric responses are fundamental to optical information processing~\cite{fan2012all,sounasNonreciprocal2017,tzuangNonreciprocal2014,yuComplete2009}, allowing extensive manipulation of light propagation. Recent developments of flat optics open up new opportunities in this direction~\cite{schulzRoadmap2024a}. For example, flat optics with electromagnetic nonlinearities have enabled optical nonreciprocity~\cite{cotrufoPassive2024, tripathiNanoscale2024}, offering unique benefits like the absence of external bias and ultra-compactness at the sub-wavelength scale. Asymmetric nonlinear frequency conversion~\cite{boroviksDemonstration2023, krukAsymmetric2022, shitritAsymmetric2018} in metasurfaces and tunable nonlinear tensor in twisted two-dimensional materials~\cite{tangOnchip2024a} were also recently realized in experiments. However, the implications of asymmetric response have been limited to the classical regime in flat optics. While various quantum phenomena arising from asymmetric response have recently received significant attention in other platforms such as optomechanical~\cite{huangNonreciprocal2018} or atomic systems~\cite{dongAlloptical2021, zhangThermalmotioninduced2018, binNonreciprocal2024a, grafNonreciprocity2022}, introducing asymmetric responses to tailor SPDC processes remained experimentally unexplored in any nonlinear systems.

Here we propose and experimentally demonstrate a novel mechanism, through the control of asymmetric nonlinear response, for generating optically tunable biphoton quantum entanglement, overcoming the tunability limit set by material nonlinear tensors. We develop a resonant semiconductor metasurface incorporating [110] InGaP nanostructures, where structural asymmetry breaks the rotational symmetry of the nonlinear polarization. The degree of symmetry breaking is controlled via the biphoton wavelength, allowing dynamic manipulation of asymmetric second harmonic generation (SHG).
Leveraging this unique characteristic, we realize continuous tuning of polarization-entangled states in a wide range, from a near-unentangled state to a Bell state via the pump wavelength. Additionally, we experimentally detect the spatial anti-correlations of photon pairs and theoretically confirm the generation of tunable hyperentanglement in the polarization and spatial degrees of freedom. This unique tuning mechanism may unlock coherent control of quantum entanglement at picosecond speeds via a pulse-shaped pump. 

Furthermore, the InGaP metasurface platform introduced here is a leading candidate for next-generation ultra-compact quantum light sources, featuring a large second-order susceptibility $\chi_{xyz}$ of 220~pm/V~\cite{uenoSecondordera}, a wide bandgap of $\sim$~650 nm and a high crystal quality due to lattice-matching to GaAs substrate in epitaxial growth. Consequently, the measured SPDC rate and coincidence-to-accidental ratio (CAR) are record-high at the telecommunication wavelength in flat-optics photon-pair sources. The achieved CAR of $\approx 7\times10^4$ exceeds those reported in other semiconductor flat optics by two orders of magnitude~\cite{guoUltrathin2023, weissflogTunable2024, santiago-cruzResonant2022, sultanovFlatoptics2022, sonPhoton2023, fengPolarizationentangled2024}. This ultra-compact and low-noise metasurface-based photon pair source with optically tunable quantum entanglement is well-suited for future monolithic integration with InGaP lasers, modulators, and detectors~\cite{liuIII2016}, offering unprecedented opportunities for quantum technologies. 

\section*{Results}
\begin{figure*}[!b] \centering
    \includegraphics[width=1\textwidth]{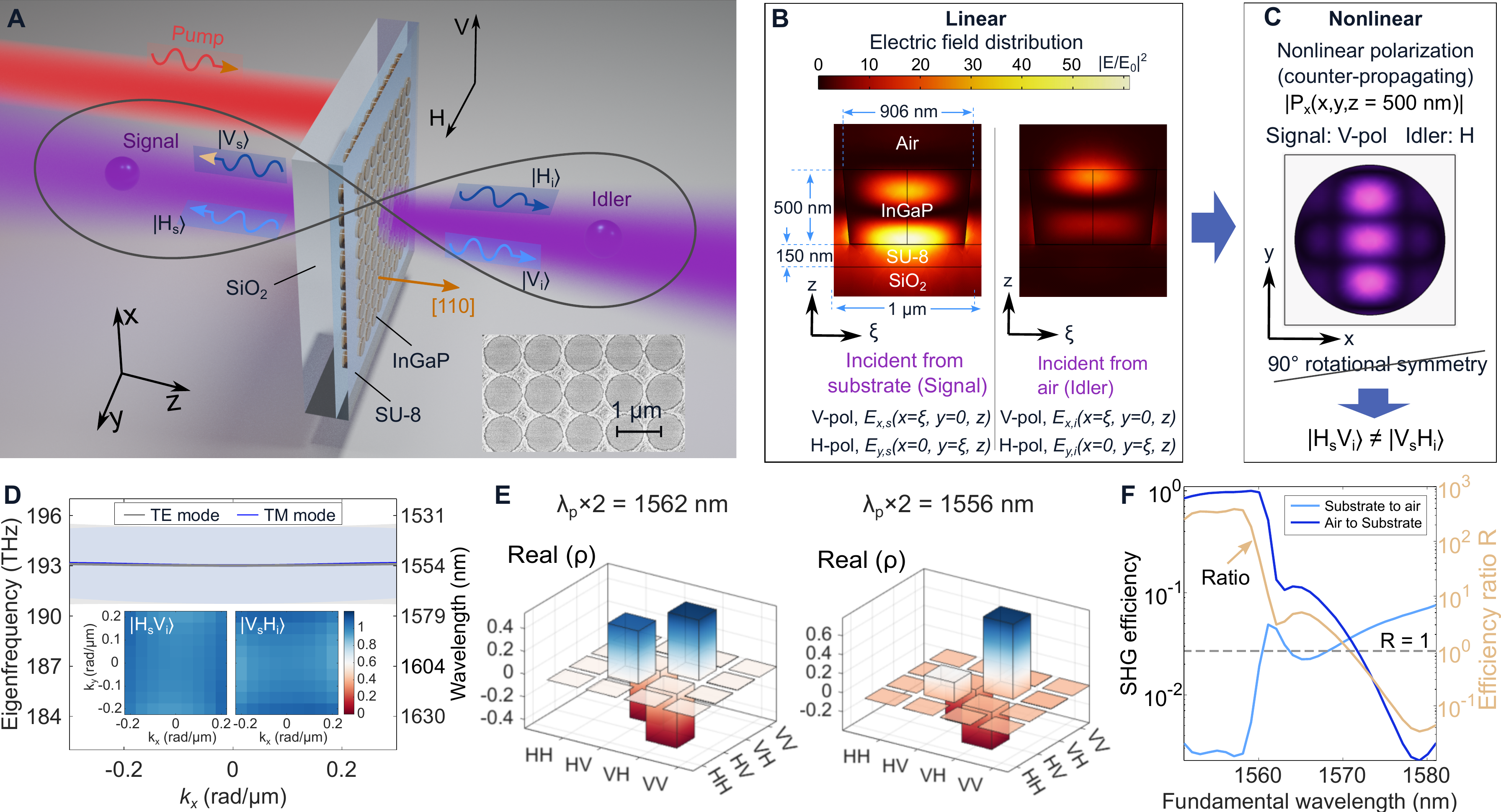} 
    \caption{\textbf{Generation of optically tunable polarization entanglement in an InGaP metasurface.} 
    (\textbf{A})~Sketch of polarization entanglement from an InGaP metasurface. The metasurface incorporates a nanostructured InGaP film with crystal orientation [110], enabling enhanced generation of polarization-entangled photon pairs counter-propagating along the $z-$direction. For convenience, we denote $-z$ and $+z$ direction as signal and idler arms, respectively. 
    The bottom right panel presents the scanning electron microscope image of the fabricated metasurface. 
    (\textbf{B}-\textbf{C})~Profiles of optical resonances and nonlinear polarization in the metasurface. The electric field within the InGaP nanostructures distributes differently as the beam is incident from the substrate (signal arm) or air side (idler arm) of the metasurface (panel \textbf{B}). This breaks the 90-degree symmetry of the nonlinear polarization (panel \textbf{C}), leading to different nonlinear processes for $\ket{H_sV_i}$ and $\ket{V_sH_i}$. The origin is located at the center of the InGaP nanopillar's bottom surface. 
    (\textbf{D})~Simulated eigenfrequency of optical resonances vs. transverse wavenumber $k_x$. The eigenfrequencies of the TE and TM modes supported by the metasurface manifest a flat band within the linewidth of optical resonances due to the localized nature of the modes. As a result, the emission patterns of SPDC processes $\ket{H_sV_i}$ (left inset) and $\ket{V_sH_i}$ (right inset) are close to flat in the $k$ space, ensuring their spatial overlaps.
    (\textbf{E})~Simulated density matrices at different pump wavelengths. At $\lambda_p\times2 = 1562$~nm, a polarization Bell state can be produced with a high fidelity of 98\% and a concurrence of 0.99. A different polarization-entangled state is produced as the pump wavelength is changed to $\lambda_p\times2 = 1556$~nm. 
    (\textbf{F})~Asymmetric SHG. Two SHG processes are simulated: excitation from the substrate (or air) and collection from the air (or substrate). Their efficiency ratio of $R = 1$ corresponds to fully symmetric SHG.
}
\label{fig:1}
\end{figure*}

\subsection*{Concept and modeling}
The nonlinear metasurface incorporates 500~nm-thick InGaP nanostructures with a [110] crystal orientation, providing a specific nonlinear polarization along the $x$-direction, i.e., $P_x = 4\epsilon_0 d_{36}(E_{x,s}E_{y,i}+E_{y,s}E_{x,i})$ (Section S1). Here $d_{36} = \frac{1}{2}\chi_{xyz}$ is the effective nonlinear coefficient and $\epsilon_0$ is the vacuum permittivity. The subscripts of the electric field $E$ denote direction with $x,y$ and signal/idler photons with $s,i$. This inherently enables the generation of cross-polarized photon pairs under an $x$-polarized pump, showing a symmetry between $E_{x,s}E_{y,i}$ and $E_{y,s}E_{x,i}$ processes.

We consider counter-propagating SPDC sketched in Fig.~\ref{fig:1}\textbf{A}, which naturally splits the signal and idler photons and allows unique entanglement tunability. 
The polarization of signal/idler photons
is in a quantum superposition of $\ket{H_sV_i}$ and $\ket{V_sH_i}$ (where $V$ and $H$ represent $x$ and $y$ directions). To ensure quantum coherence, a high spatial overlap between $\ket{H_sV_i}$ and $\ket{V_sH_i}$ is a prerequisite. This is achieved by tailoring metasurface optical resonances to be localized (see Fig.~\ref{fig:1}\textbf{B}), where the resonance wavelengths ($\approx$1554~nm) for TE and TM modes form nearly a flat band within their linewidths across a large range in $k$-space (see Fig.~\ref{fig:1}\textbf{D}). The flat band is enabled by the high refractive index $\sim 3.12$~\cite{uenoSecondordera} of InGaP.
Consequently, the simulated biphoton spatial emission rate for both $\ket{H_sV_i}$ and $\ket{V_sH_i}$ processes are flat and manifest high overlap in $k$-space (see insets), enabling high-quality polarization entanglement. As shown in Fig.~\ref{fig:1}\textbf{E}, the simulated density matrix reveals an entangled polarization Bell state $\ket{\Psi^-} = \frac{1}{\sqrt{2}}(\ket{H_sV_i}-\ket{V_sH_i})$ at the pump wavelength $\lambda_p\times2 = 1562$~nm, with a fidelity $>98\%$ and a concurrence $>0.99$. The minus sign in $\ket{\Psi^-}$ is caused by a $\pi$ phase difference in counter-propagating SPDC processes. Notably, a differently polarization-entangled state is generated at a different pump wavelength $\lambda_p\times2 = 1556$~nm. The control of polarization entanglement is enabled by a new tuning mechanism --- the symmetry breaking of the nonlinear polarization within the nanostructures. 

To understand this mechanism, we analyze classical nonlinear processes SHG and sum frequency generation (SFG), which also reveal the information about SPDC.
Figure~\ref{fig:1}\textbf{B} reveals that co-polarized signal/idler photons incident from the substrate or air manifest different electric field distributions inside the InGaP in the $x$-$y$ plane for a given $z$, i.e., $E_{x,s}^{(V_s)}(x,y) \neq E_{x,i}^{(V_i)}(x,y)$ and $E_{x,s}^{(H_s)}(x,y) \neq E_{x,i}^{(H_i)}(x,y)$, where the superscripts represent the photon polarization. This characteristic is caused by the geometric asymmetry of the sample including the metasurface and a substrate in the $z$-direction. 
As a result, the SHG manifests a diode-like asymmetric response~\cite{boroviksDemonstration2023, shitritAsymmetric2018} controllable via the fundamental wavelength (see Fig.~\ref{fig:1}\textbf{F}). Meanwhile, the nonlinear polarization for counter-propagating 
signal/idler photons breaks the $90^\circ$ rotational symmetry in the $x$-$y$ plane (see Fig.~\ref{fig:1}\textbf{C}), 
leading to asymmetric sum frequency generation (SFG) processes for $P_{x}^{(V_{s}H_{i})}(x,y)$ and $P_{x}^{(H_{s}V_{i})}(x,y)$. 
Due to the classical-to-quantum correspondence~\cite{marinoSpontaneous2019, parryEnhanced2021}, the quantum efficiency of $\ket{H_sV_i}$ and $\ket{V_sH_i}$ processes can also differ.
As the pump frequency changes, the signal and idler wavelengths vary along with their respective mode profiles within the nanostructures. This alters the symmetry of the corresponding nonlinear polarization and thus the tunning of polarization entanglement. Note that such tunability induced by symmetry-breaking is not applicable for co-propagating SPDC in the present metasurface, due to the $C_{4v}$ symmetry of the nanostructures.
A more detailed and quantitative discussion is provided in Section S2 of the supplement.

\subsection*{Experimental characterization of enhanced nonlinear processes}

\begin{figure*}[!b] \centering
\includegraphics[width=0.94\textwidth]{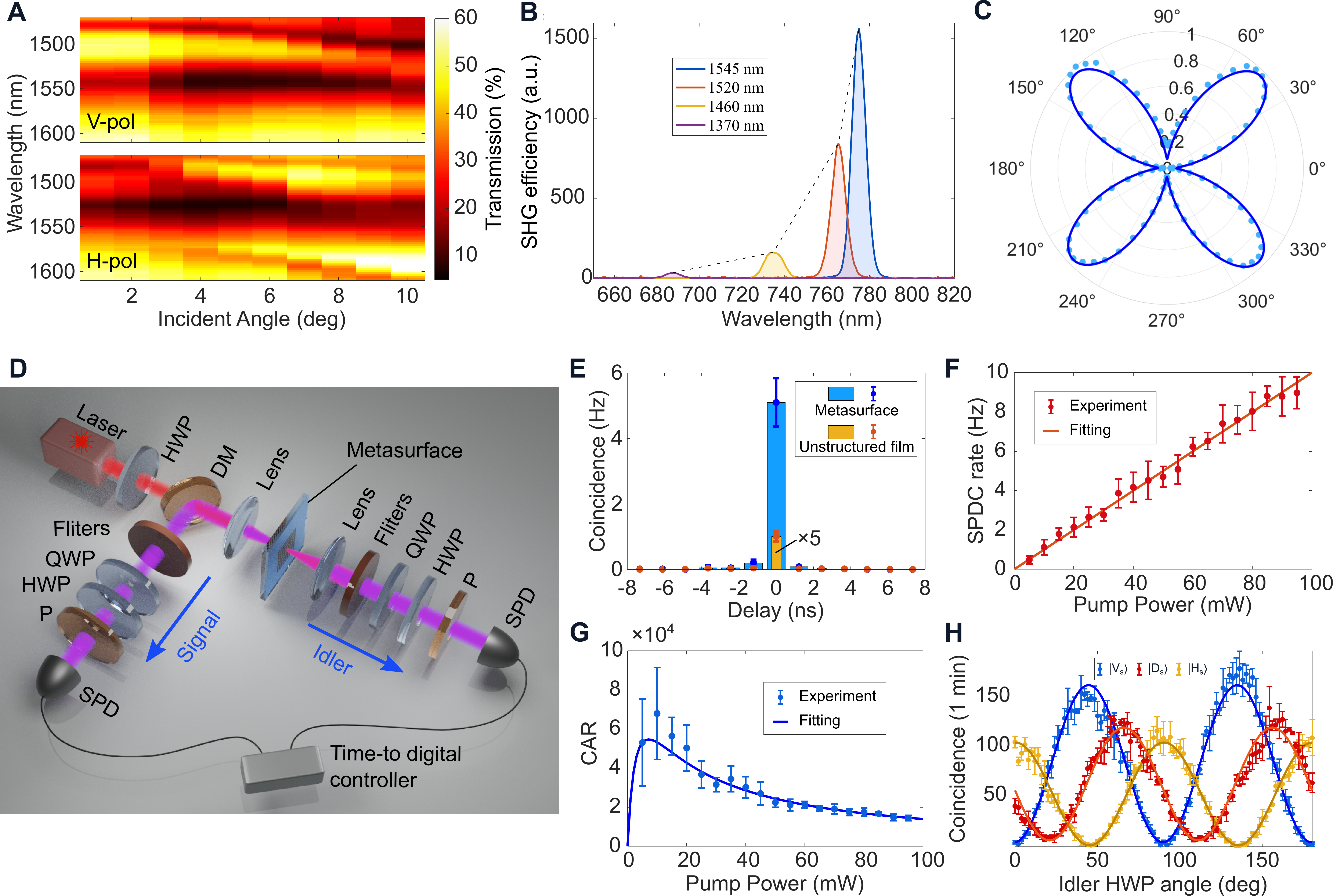} 
\caption{\textbf{Experimental characterization of nonlinear processes.} 
    (\textbf{A})~Measured linear transmission spectra vs. incident angle for V- and H- polarized beams. The fabricated metasurface manifests optical resonances at $\sim$1550~nm with a flat band for both polarizations. 
    (\textbf{B})~Enhanced generation of SHG. The SHG efficiency is enhanced by 50 times at resonance wavelength (blue curve) as compared to off-resonance excitation (purple curve). The maxima are connected with a dashed line to enhance clarity.
    (\textbf{C})~SHG efficiency vs. polarization of the fundamental beam. The SHG efficiency is maximized at $D$-polarized fundamental beam, attributing to the nonlinear tensor of InGaP [110]. 
    (\textbf{D})~Experimental setup for characterizing SPDC. The back-propagating signal photons reflected by the dichroic mirror (DM) and the forward-propagating idler photons go through a set of optical filters, after which their polarizations are analyzed with a half-wave plate (HWP), quarter-wave plate (QWP), and a polarizer (P). 
    (\textbf{E})~Coincidence histograms for metasurface and unstructured thin film. The measured photon pair rate from the metasurface reveals a 25-time enhancement compared to that from the unstructured thin film. 
    (\textbf{F})~SPDC rate vs. pump power. The measured SPDC rate linearly depends on the pump power. 
    (\textbf{G})~CAR vs. pump power. The correlation measurement manifests extremely high CAR values, reaching a maximum of $\approx 7\times 10^4$. 
    (\textbf{H}) Coincidence counts vs. HWP plate angle at idler arm $\theta_{H,i}$. We fix polarization projections of signal photons to $\ket{V_s}$, $\ket{D_s}$, $\ket{H_s}$, respectively. The visibility for these three cases is $>89\%$, violating the Bell equality and confirming the presence of entanglement. Error bars indicate one standard deviation (SD) in all plots.}
\label{fig:2}
\end{figure*}

We first experimentally measure the angle-dependent linear transmission of the metasurface
(see Fig.~\ref{fig:2}\textbf{A}), where optical resonances are signified by transmission minima for H- and V-polarized inputs overlap at $\approx$~1550~nm with $Q\approx30$. Both modes manifest a flat band across a large incident angle range of $\pm 10$ degrees. 
To understand the optical enhancement and nonlinear tensor, 
we characterize the V-polarized SHG from the InGaP metasurface, as shown in Fig.~\ref{fig:2}\textbf{B-C}. The SHG is enhanced 50 times at the resonance wavelength compared to the off-resonance excitation. 
The SHG efficiency is maximized for a diagonal or anti-diagonal polarized fundamental beam, confirming the presence of the nonlinear tensor component $P_x$. 


We proceed with SPDC characterization using the experimental setup shown in Fig.~\ref{fig:2}\textbf{D}, where detailed procedures are provided in Section S4 of the supplement. With the same collection angle of $3^\circ$, the measured coincidence rates of the photon pairs from the metasurface and unstructured InGaP film are compared in Fig.~\ref{fig:2}\textbf{E}, revealing a 25-fold SPDC enhancement due to the optical resonances supported by the metasurface. We also find that the measured SPDC rate depends linearly on the pump power (see Fig.~\ref{fig:2}\textbf{F}). Linear fitting to experimental data yields a detected SPDC rate of 0.1~Hz/mW, which is record-high for flat optics quantum light sources at the telecommunication wavelengths (see Section S5). This high rate is attributed to the large second-order nonlinear coefficient, the use of $[110]$ crystal orientation, and flat biphoton emission.

We present the CAR of biphoton emission from the metasurface at different pump powers in Fig.~\ref{fig:2}\textbf{G}. The solid curve is obtained from linear fitting of the SPDC rate and single-photon count rates (Section S5). The decrease in CAR at pump powers $<5$ mW is due to dark counts. Notably, the maximum CAR value reaches $\approx 7\times10^4$, which is record-higher than in any existing photon-pair source with flat optics (see the benchmarking in Section S5). This CAR exceeds those reported in previous works on semiconductor flat optics by three orders of magnitude, including two-dimensional materials~\cite{guoUltrathin2023, weissflogTunable2024} and other III-V semiconductors~\cite{santiago-cruzResonant2022, sultanovFlatoptics2022}. The CAR also exceeds those in metasurfaces~\cite{zhangSpatially2022, santiago-cruzPhoton2021} incorporating dielectric material traditionally used for SPDC. 
This breakthrough is primarily attributed to the wide bandgap of InGaP at 1.9~eV (650~nm) and the high crystal quality attained through epitaxial crystal growth (see Section S3 for detailed material characterization), resulting in minimal fluorescence at $\lambda_p>650$~nm. Additionally, the high second-order nonlinearity of 220 pm/V~\cite{uenoSecondordera} and optical resonances supported by the metasurface boost the SPDC process without adding more noise. 
With such ultra-low noise, the SPDC rate can be significantly increased by enhancing pump power through additional optical resonance at the pump wavelength while maintaining a high CAR.
These results suggest that the InGaP metasurface platform is highly attractive for future SPDC-based quantum technologies.

We analyze the polarization properties of the generated photon pairs by varying the idler HWP angle $\theta_{H,i}$ for three signal polarization projections $\ket{V_s}$, $\ket{D_s}$, and $\ket{H_s}$. 
Figure.~\ref{fig:2}\textbf{H} shows periodic oscillations in all three cases.
The smallest visibility is $\nu = 89\pm2$\% for the $\ket{D_s}$ projection, 
which is larger than 71\%. This violates the Bell inequality by 9 SD, confirming the generation of polarization entanglement~\cite{JinPulsed2014,clauserBell1978}.

\begin{figure*}[!b] \centering
    \includegraphics[width=0.89\textwidth]{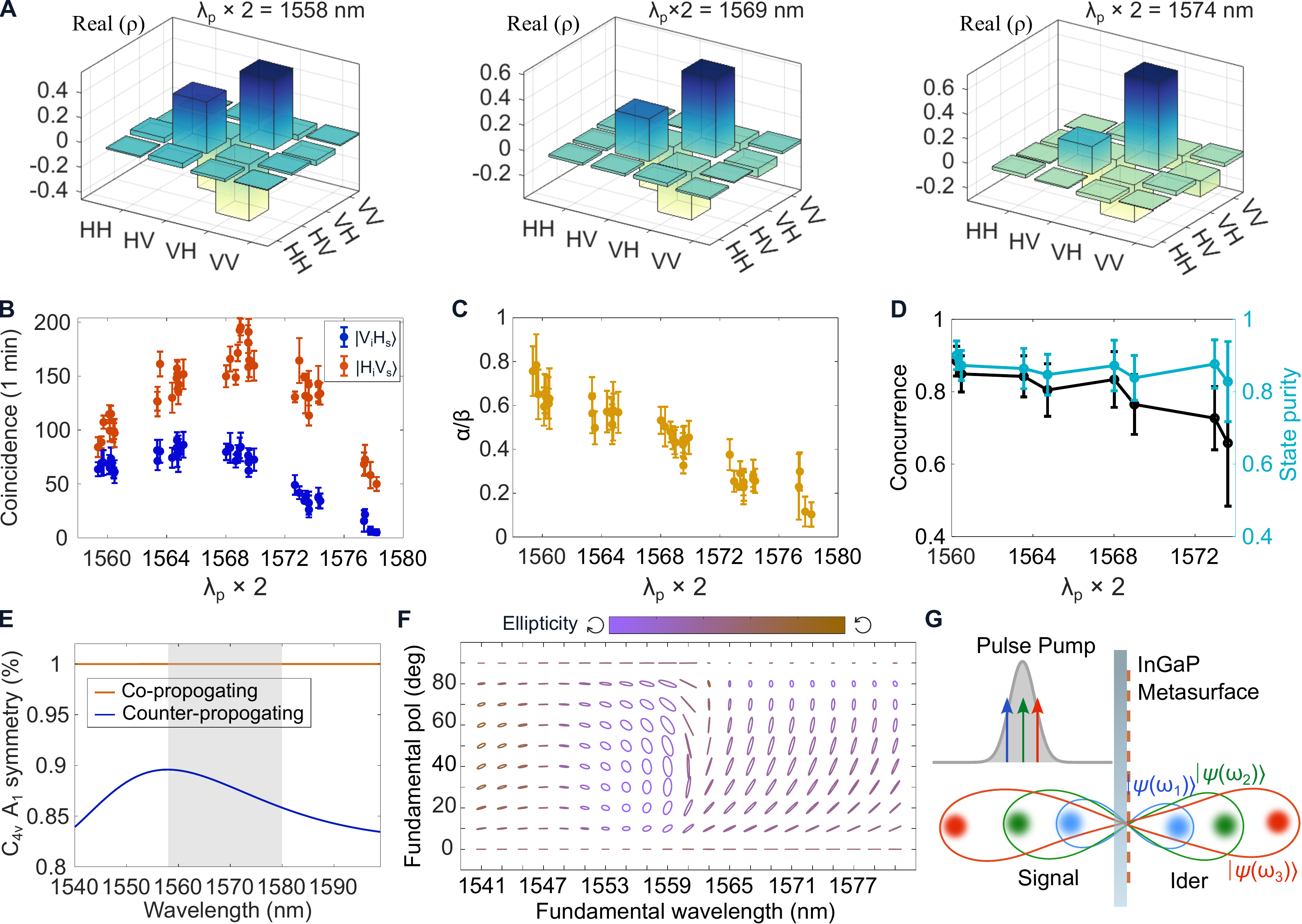} 
    \caption{\textbf{Experimental characterization of optically tunable polarization entanglement.} (\textbf{A})~Reconstructed density matrix at pump wavelength $\lambda_p\times2 = 1558$~nm. The density matrix presents polarization entanglement close to a fully entangled Bell state $\ket{\Psi^-} = \frac{1}{\sqrt{2}}(\ket{H_sV_i}-\ket{V_sH_i})$, with fidelity $>95$\% and concurrence $>0.9$. The density matrix changes at different pump wavelengths $\lambda_p\times2 = 1569$~nm or $1574$ nm. 
    (\textbf{B}) Measured photon-pair rates vs. pump wavelength. The biphoton coincidences measured at polarization projections $\ket{H_sV_i}$ and $\ket{V_sH_i}$, respectively show distinct dependence over pump wavelengths. 
    (\textbf{C})~Coefficient $\alpha/\beta$ of polarization-entangled state vs. pump wavelength. The coefficient $\alpha/\beta$ calibrated from panel \textbf{C} is optically tunable by changing the pump wavelength. 
    (\textbf{D})~Concurrence and state purity vs. pump wavelength. The concurrence of the reconstructed density matrix drops as the pump wavelength is tuned while the state purity remains high. 
    (\textbf{E}) Symmetry analysis of the nonlinear polarization by projecting its field onto the $A_1$ symmetry of the $C_{4v}$ point group. The degrees of symmetry breaking enable the tuning of the state coefficients. The shaded region represents the range of $\lambda_p\times2$ used in panels \textbf{C-D}. 
    (\textbf{F})~Simulated polarization of counter-propagatingSHG vs. the fundamental wavelength and polarization, indicating extensive tunability of entanglement. Each ellipse represents the SHG polarization state in phase space, with its size indicating the SHG efficiency and its color suggesting ellipticity.
    (\textbf{G})~Perspective with a pulsed-shaped pump. Pumping the metasurface with a shaped pulse allows ultra-fast tuning of polarization entanglement, enabling high-speed information encoding. Error bars indicate one SD in all plots.}
\label{fig:3}
\end{figure*}

\subsection*{Experimental confirmation of optically tunable polarization entanglement}

Next, we experimentally perform full quantum tomography and reconstruct density matrices using the Maximum Likelihood Method~\cite{jamesMeasurement2001}. All following measurements shown in Fig.~\ref{fig:3}\textbf{A} are performed using a V-polarized pump, with results for an H-polarized pump provided in Section S6 of the supplement.
At $\lambda_p\times 2 = 1558$~nm, the reconstructed density matrix reveals a polarized Bell state $\ket{\Psi^-} = \frac{1}{\sqrt{2}}(\ket{H_sV_i}-\ket{V_sH_i})$ with fidelity of 95\% and concurrence of 0.9. Most importantly, the polarization entanglement can be tuned by varying the pump wavelength. For example, at $\lambda_p\times2 = 1574$~nm, the $\ket{V_sH_i}$ projection becomes more dominant over $\ket{H_sV_i}$. 

To explore such tunability further, we measure the SPDC rate under $\ket{V_sH_i}$ and $\ket{H_sV_i}$ projections versus $\lambda_p\times2$ as shown in Fig.~\ref{fig:3}\textbf{B}, where the unevenly spaced wavelengths are due to laser diode mode-hopping.
We observe differing wavelength dependence in measured coincidences for $\ket{V_sH_i}$ and $\ket{H_sV_i}$, indicating that each pump wavelength corresponds to a different polarization-entangled state $\rho$ with distinct diagonal coefficients  $\alpha = \mel{H_sV_i}{\rho}{H_sV_i}$ and $\beta = \mel{V_sH_i}{\rho}{V_sH_i}$. The measured coincidence rates of $\ket{H_sV_i}$ and $\ket{V_sH_i}$ are proportional to coefficients $\alpha$ and $\beta$ respectively, with their ratio $\alpha/\beta$ shown in Fig.~\ref{fig:3}\textbf{C}. The ratio can be controlled from 0.1 (a near-unentangled state) to 0.8 (approaching a maximally entangled Bell state) by simply changing $\lambda_p$; meanwhile the SPDC rate is enhanced as compared to unpatterned thin films across these pump wavelengths. 
Furthermore, we characterize the concurrence and purity of the reconstructed quantum states (see Fig.~\ref{fig:3}\textbf{D}). The concurrence decreases at longer pump wavelengths from $\sim$0.9 to $\sim$0.7 because the quantum state changes from a Bell state to a less entangled state. However, the state purity remains above $\sim$0.8 for all pump wavelengths, suggesting that all entangled states produced here are indeed in a coherent superposition of $\ket{H_sV_i}$ and $\ket{V_sH_i}$. 

To provide a physical insight into the measurement results, we perform a quantitative symmetry analysis~\cite{Parry2024} of the nonlinear polarization by projecting its distribution onto the irreducible representations of the $C_{4v}$ point group. We consider 20 $x$-$y$ planes equally distributed along the $z$-axis within the nanostructures and analyze the overall symmetry across each plane weighted by the average amplitude of the in-plane nonlinear polarization. The $A_1$ symmetry parameter $\eta$ versus pump wavelength, indicating the degree of symmetry (ranging from 0 to 1), is shown in Fig.~\ref{fig:3}\textbf{E} (see detailed analysis in Section S2). In the co-propagating nonlinear process, the nonlinear polarization maintains a high degree of symmetry ($\eta \sim 1$) across all wavelengths. However, for the counter-propagating case, the dependence of $\eta$ on the pump wavelength correlates with changes in quantum states. At $\alpha/\beta \approx 1$, $\eta$ tends to be maximized, suggesting the largest symmetry. The ratio $\alpha/\beta$ decreases or increases as $\eta$ becomes smaller, thereby breaking the symmetry of nonlinear polarization. 
To our knowledge, this is the first quantum metasurface manifesting an asymmetric nonlinear response. 

Furthermore, a more comprehensive entanglement tunability can be achieved by including pump polarization as an additional tuning parameter. The simulated SHG for counter-propagating signal/idler, presented in Figure~\ref{fig:3}\textbf{F}, shows that fundamental wavelength and polarization provide extensive control over SHG polarization. The use of different optical modes also reveals the control in distinct regimes (Section S7). These findings indicate a significant potential to broadly extend the range of entanglement tunability. While pump polarization tuning is generally slow as previously discussed, its combination with wavelength-enabled ultrafast tuning may facilitate dynamic tuning of on-demand quantum states unattainable with unstructured nonlinear crystals.

This novel tunability can potentially stimulate a wide range of applications with a pulse-shaped pump (see Fig.~\ref{fig:3}\textbf{G}) whose Fourier transform presents a frequency comb~\cite{reimerGeneration2016a, gaetaPhotonicchipbased2019, pasquaziMicrocombs2018}.  
Each tooth of the comb carries one pump frequency for generating a unique polarization-entangled state at the corresponding biphoton wavelength. A direct application is multiplexing polarization entanglement into each wavelength channel~\cite{wengerowskyEntanglementbased2018a, herbautsDemonstration2013}
, allowing the distribution of quantum states to multiple users. A key benefit of our approach is that the entanglement can be tailored in each channel for independent data transmission to different users. 
Additionally, time-division multiplexing can be achieved by allowing one photon to travel through a dispersive medium such that its travel time varies depending on its wavelength, while the other photon propagates through free space. As a result, each entangled state is encoded into different time bins at picosecond resolution according to its wavelength, enabling ultrafast switching of polarization entanglement. These multiplexing schemes may allow multi-user networks~\cite{wengerowskyEntanglementbased2018a, herbautsDemonstration2013, donohueUltrafast2014}, and the ultrafast switching may be suited for routers compatible with existing high-speed telecommunication infrastructure~\cite{hallUltrafast2011}. 
Furthermore, this tuning approach may enable the generation of tunable multiphoton states. For example, we consider two independent biphoton entanglement generated from two teeth of the frequency comb, $\ket{\psi_1} = a_1\ket{H_{s,1}V_{s,1}}+b_1\ket{V_{s,1}H_{s,1}}$ and $\ket{\psi_2} = a_2\ket{H_{s,2}V_{s,2}}+b_2\ket{V_{s,2}H_{s,2}}$. By post-selecting four-photon events~\cite{reimerGeneration2016b}, the generated state becomes $\ket{\psi} = \ket{\psi_1}\otimes\ket{\psi_2} = a_1a_2\ket{H_{s,1}V_{s,1}H_{s,2}V_{s,2}}+a_1b_2\ket{H_{s,1}V_{s,1}V_{s,2}H_{s,2}}+b_1a_2\ket{V_{s,1}H_{s,1}H_{s,2}V_{s,2}}+a_2b_2\ket{V_{s,1}H_{s,1}V_{s,2}H_{s,2}}$, where the state coefficients can be controlled via pump wavelength and polarization. In these applications, our approach allows real-time tuning and is scalable depending on the number of teeth in the pump frequency comb.

\begin{figure*}[!b] \centering
    \includegraphics[width=1\textwidth]{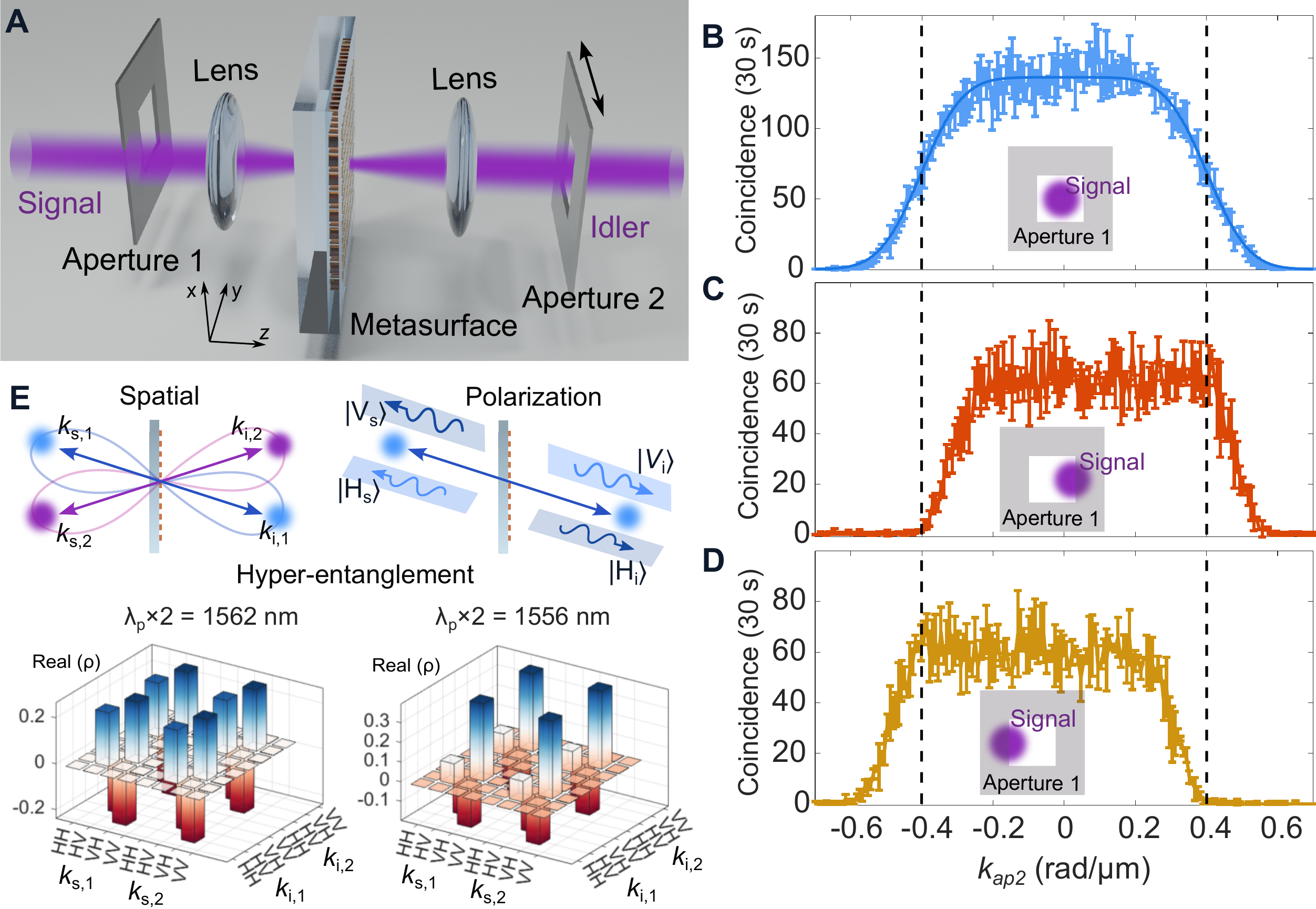} 
    \caption{\textbf{Perspective of hyperentanglement.} 
    (\textbf{A})~Experimental setup for measuring spatial correlations. 
    Aperture~1 is fixed at different positions $k_{\rm ap,s}$, while aperture~2 position $k_{\rm ap,i}$ is scanned through the idler emission along the $y$-axis. The origins of $k_{\rm ap,s}$ and $k_{\rm ap,i}$ are respectively defined as the central position of apertures aligning with the signal/idler emission center.
    (\textbf{B,C,D})~Coincidence vs. position of aperture~2 $k_{\rm ap,i}$. The position of aperture~1 is set as $k_{\rm ap,s} = 0,-L/2$ and $L/2$ respectively. When the right half of the signal emission is blocked (panel \textbf{C}), coincidences collected at $k_{\rm ap,i} = L/2$ are closed to the maximum, and almost no idler photons are detected at $k_{\rm ap,i} = -L/2$. Similar behaviorr is observed when the left half of the signal emission is blocked (panel \textbf{D}), indicating the anti-spatial correlations of signal and idler photons and the superposition of different spatial modes. The error bars indicate one SD in all plots. 
    (\textbf{E}) Simulated density matrix of hyperentanglement. As shown in the sketch, the photon pairs from the metasurface can manifest spatial ($\ket{\psi_{\rm spatial}}= \frac{1}{\sqrt{2}}\ket{k_{s,1}k_{i,1}}+\ket{k_{s,2}k_{i,2}}$) and polarization ($\ket{\psi_{\rm pol}} = a_p\ket{H_sV_i}+b_p\ket{V_sH_i}$) entanglements. The simulated density matrix, including both spatial and polarization degrees of freedom, suggests the presence of hyperentanglement, $\ket{\Psi_{\rm hyper}} = \ket{\psi_{\rm spatial}}\otimes\ket{\psi_{\rm pol}}$. The hyperentanglement is optically tunable by changing the pump wavelength.}
\label{fig:4}
\end{figure*}

\subsection*{Experimental characterization of spatial correlations and perspective in hyperentanglement}

This new tuning mechanism can be extended to control photon pairs hyper-entangled in more than one degree of freedom, which can bring unlimited possibilities to future quantum technologies. 
We first calibrate the spatial correlations of signal and idler photons in addition to their polarization, using the setup illustrated in Fig.~\ref{fig:4}\textbf{A}. We remove all waveplates and polarizers and add a square aperture with width $L$ in the Fourier plane ($k$-space) for both signal and idler arms. At $k_{\rm ap,s}=0$ (see Fig.~\ref{fig:4}\textbf{B}), coincidences measured against $k_{\rm ap,i}$ provide spatial information about photon-pair emission and collection (see Section S8). Figures~\ref{fig:4}\textbf{C-D} present spatial correlation measurements at $k_{\rm ap,s} = -L/2$ and $L/2$. Almost no photon pairs are detected when $k_{\rm ap,s} = k_{\rm ap,i} = -L/2$ or $L/2$; meanwhile maximum coincidences occur at $k_{\rm ap,s} = -k_{\rm ap,i} = L/2$ and $-L/2$. These observations prove robust spatial anti-correlation of photon pairs in $k$-space with opposite transverse momenta. Additionally, the simultaneous presence of $\ket{k_{\rm ap,s} = L/2, k_{\rm ap,i} = -L/2}$ and $\ket{k_{\rm ap,s} = -L/2, k_{\rm ap,i} = L/2}$ indicates the superposition of two spatial quantum states and spatial entanglement. 
The complete characterization of entanglement requires additional measurements at different spatial projections~\cite{setzpfandtTunable2016b}, which goes beyond the scope of this work but presents an interesting avenue for future studies.
Note that the polarization entanglement is reconstructed by summing over all spatial modes and spatial correlations are measured without polarization projections, suggesting the presence of a hyper-entangled state being a product state of polarization and spatial states. 

Inspired by the experiment, we theoretically confirm the generation of hyperentanglement from the metasurface (see Fig.~\ref{fig:4}\textbf{E}). As a proof of principle, we analyze the spatial entanglement $\ket{\psi_{\rm spatial}}= \frac{1}{\sqrt{2}}(\ket{k_{s,1}k_{i,1}}+\ket{k_{s,2}k_{i,2}})$ of two spatial modes, where $k_{s,1} = -k_{i,1}$ and $k_{s,2} = -k_{i,2}$. Under each spatial mode projection, polarization entanglement $\ket{\psi_{\rm pol}} = a_p\ket{H_sV_i}+b_p\ket{V_sH_i}$ is present. Consequently, a hyper-entangled state $\ket{\Psi_{\rm hyper}} = \ket{\psi_{\rm spatial}}\otimes\ket{\psi_{\rm pol}}$ can be generated, as reflected by the simulated density matrices. 
The predicted fidelity of quantum state produced at $\lambda_p\times2 = 1562$~nm to the ideal hyper-entangled state $\ket{\Psi_{\rm hyper}}$ yields 98\%, indicating the practical feasibility of generating high-quality hyperentanglement. Furthermore, the hyperentanglement can also be optically controlled via pump wavelength through symmetry breaking. A significant advantage of metasurface-generated hyperentanglement is that the number of allowable spatial modes can be $>600$ times higher than that from bulky crystals (see Section S8). This is due to the constraints imposed by longitudinal phase matching conditions in bulky crystals, which limits the biphoton emission range in transverse $k$-space. The use of a small pump beam waist due to a limited aperture of bulky crystals further restricts the spatial bandwidth of each mode.

\section*{Discussion}
This work uncovers a new mechanism, by breaking the symmetry of nonlinear polarization, for producing optically tunable quantum entanglement in a nonlinear semiconductor metasurface, ranging from polarization Bell state to hyperentanglement.  We show that the number of spatial modes in hyperentanglement can potentially outperform bulky crystals by more than three orders of magnitude.
The fully entangled polarization Bell state demonstrated is naturally compatible with established quantum communication protocols~\cite{liaoSatellitetoground2017}, while the partially entangled state can be used for Bell inequality tests~\cite{giustinaFree2015}, quantum teleportation~\cite{modlawskaNonmaximally2008}, and quantum key distribution~\cite{xueEfficient2001}. Furthermore, this tuning mechanism may unlock a wide range of novel applications when combined with a pulse-shaped pump containing a frequency comb, such as wavelength-division multiplexing~\cite{wengerowskyEntanglementbased2018a, herbautsDemonstration2013}, time-division multiplexing~\cite{tangMeasurementDeviceIndependent2016, frohlichQuantum2013}, ultrafast entanglement switching~\cite{donohueUltrafast2014}, tunable multiphoton states~\cite{reimerGeneration2016b}, and more. 


Introducing asymmetric nonlinear responses in quantum flat optics can open new directions to tailor on-demand quantum states by engineering the metasurface geometry. This approach builds on the feasibility demonstrated in previous studies of classical asymmetric nonlinear processes~\cite{boroviksDemonstration2023, krukAsymmetric2022}. Leveraging symmetry breaking, the tuning parameter could potentially be extended to include spatial, spectral, and polarization degrees of freedom. Beyond controlling polarization entanglement, this approach may be developed to tailor spatial, spectral, or time-bin entangled states, as well as their hyperentanglement.


Furthermore, our InGaP nanostructures outperform other flat-optics photon pair sources in many aspects. Their ultra-high nonlinearity and low fluorescence enable a record high SPDC rate and CAR in infrared flat-optics quantum light sources.
Additionally, this platform may be suited for the monolithic integration of III-V semiconductor lasers, modulators, and detectors on a single chip, leading to ultra-compact, robust, and multifunctional quantum photonic devices. 

Although the photon-pair rate from the InGaP metasurface is currently lower than that of traditional bulky nonlinear crystals, new techniques such as triple resonances at pump, signal, and idler wavelengths are anticipated to boost the SPDC rate significantly. The InGaP metasurface with exceptional control over quantum states can be potentially combined with emerging InGaP integrated photonics platforms (such as microrings~\cite{zhaoInGaP2022b} or waveguides~\cite{akinInGaP2024a}) with an SPDC rate 
outperforming traditional nonlinear material platforms. Combining the unique functionalities demonstrated here, these promising developments position InGaP metasurfaces as attractive candidates for next-generation quantum light sources, offering unprecedented opportunities for quantum information processing.

\bibliography{mybib2}

\bibliographystyle{lpr}

\section*{Acknowledgments}
This work was supported by the Australian Research Council (DP190101559, CE200100010) and Meta Active IRTG 2675 (437527638). Metasurface fabrication was performed at the Australian National University node of the Australian National Fabrication Facility (ANFF), a company established under the National Collaborative Research Infrastructure Strategy to provide nano- and microfabrication facilities for Australian researchers. The authors acknowledge Sebastian Klimmer for building the automated setup for SHG measurements.

\section*{Author contributions}
All authors contributed to the experiment, the analysis of the results, and the writing of the manuscript.

\section*{Competing interests}
The authors declare no competing interests.

\section*{Data availability} 
All other data needed to evaluate the conclusions in the paper are present in the main text or the supplementary materials.

\end{document}